# Fractional size dependence of surface tension of a growing nucleus with an inhomogeneous interface

Puja Banerjee and Biman Bagchi*

Solid State and Structural Chemistry Unit, Indian Institute of Science, Bangalore, India

Corresponding author's E-mail: *profbiman@gmail.com/ bbagchi@iisc.ac.in*

## *Abstract*

*The surface energy of the nucleus of a stable phase growing in the presence of several amorphous metastable phases of character intermediate between the initial and the final phases may depend non-trivially on the size of the nucleus. This size dependence is being increasingly used to explain diverse non-equilibrium phase selection, and relaxation, as in the random first-order transition (RFOT) theory of glasses. Here we develop an order parameter based Ginzburg-Landau approach that explicitly includes the rugged free energy landscape due to the metastable phases. The fractional dependence of total surface energy between melt and stable solid phase on the number of metastable phases($N_{MS}$) has been interrogated in this study. We have also analyzed how this fractional dependence gets modified with temperature. We find that the fractional size dependence of surface energy is **omnipresent, arises from the minimization of free energy and demands certain ordering of metastable phases in the interface**. Our results could recover the celebrated result of Villain that forms the basis of RFOT theory of glasses. We find the additional result that the surface tension saturates to a finite size independent value.*

_________________________________________________________________________

Synthesis of complex solids presents several paradoxes that continue to attract attention. Zeolites present a good example. Here quartz is the stable thermodynamic phase and faujasite is one of the least stable metastable phases at low temperature. Yet, quartz precipitates out from the melt sodium alumino silicate at high temperature while faujasite forms at low temperature[1]. At intermediate temperatures one can detect the formation of other zeolite phases that of intermediate stability or metastability. In addition to zeolites, there are a large number of polymorphic systems in nature, such as phosphates[2], titanates, carbonates, silicates[1, 3 4 5] etc. The complexity of the synthesis of these inorganic and several organic solids is often due to the presence of several metastable phases. Recently, Tanaka and coworkers have shown that homogeneous nucleation of ice involves a metastable phase which they named as Ice $0$[6]. Unfortunately, there exists no satisfactory quantitative theory that of nucleation and growth includes the effects of the metastable phases of order intermediate between the melt and the growing stable phase.

The purpose of the present work is to partly remove this lacuna. We develop a statistical mechanical approach to address the effects not only of the relative depths and positions of metastable minima but also of the relative curvatures which are ignored in the classical nucleation theory and thus it fails to describe the preferential appearance of certain solids in certain temperature (and pressure) windows[7]. The theory developed here is shown to have relevance in wide range of natural phenomena, including glass transition [8-12].

The formation and precipitation of phases is often dictated by the details of free energy landscape [13]. According to classical nucleation theory, polymorph selection in these systems depends mainly on the two factors: the free energy difference between the parent and the daughter phases ($\Delta G_V$) and the surface tension between them ($\gamma$).

$$\Delta G(R) = -\frac{4\pi}{3} R^3 \Delta G_v + 4\pi R^2 \gamma \qquad (1)$$

Here R is the radius of nucleus, the only order parameter used in CNT. However, the competition between the size of the nucleus and the free energy gap ($\Delta G_V$) may be harnessed to create new metastable phases, as the free energy gap can be changed by external conditions like temperature (zeolites), pressure (core-shell systems)[14]. The second contribution from surface tension to polymorph selection is of huge interest as it can get modified depending on the morphology of metastable phases, their geometry and energetics, and thermodynamic conditions[12, 15]. A number of earlier researches suggested that surface tension, $\gamma$ should be a function of R. Long ago Gibbs[16] and Tolman[17] derived expression for radius (or, curvature) dependence of surface tension, $\gamma(R)$. However, the dependence on R predicted was weak.

The non-trivial radius dependence of the surface tension term arises from the wetting of the interface of the nucleus of the stable phase by the metastable phases. In a previous work, we showed how the presence of metastable phases helps in reducing the surface tension between bulk and stable solid phase [18]. As the size of the nucleus starts growing, it can accommodate more and more metastable phases around it. This model is popularly known as the "core-shell" model of nucleation. Although a number of earlier experimental [19-22], theoretical[12, 23] and computational[15, 24-26] studies suggested this surface wetting picture, no microscopic study has yet been carried out. In the random-field Ising model (RFIM), Villain calculated interfacial tension between spin-up region and spin-down region applying renormalization group approach and derived the scaling of surface tension with radius, R that is different from conventional, $R^{(d-1)}$ for a $d$-dimensional system[8].

$$\gamma(R) \approx \gamma_0 (R/R_0)^{-(d-2)/2} \qquad (2)$$

where $\gamma_0$ is the surface-tension coefficient at molecular length scale and $R_0$ is the interparticle spacing. Villain considered a ferromagnetic Ising model in a weak random field and the flipping of spins results in a reorganization of domain structure. The surface tension of this domain wall is calculated by considering the formation of a bump of radius, R and height, $\xi$ by domain walls in the RFIM system. The corresponding free-energy gain ($\delta G_1$) and energy loss ($\delta G_2$) can be written as $\delta G_1 \approx -H R^{d/2} \left(\frac{\xi}{R}\right)^{1/2}$ and $\delta G_2 \approx \gamma R^{d-1} \left(\frac{\xi}{R}\right)^2$, where $\gamma$ is the surface tension and H is a positive constant. By minimizing total free energy with respect to $\xi$ ($\xi<R$), one obtains the height of the bump. Then he applied renomalisation group approach to obtain the final scaling relation (Eq. (2)). Depending on the domain structure, it can be classified as stable or metastable phases in this RFIM system. According to Villain relation (Eq. (2)) as the radius increases one stable domain can be surrounded by a number of metastable domains. Kirkpatrick *et al.*[9, 27] and *Xia et al.*[10, 28, 29] utilized this relation extensively in their study of random first order transition theory of glass, where the reduction of surface tension in these studies was argued to be due to the wetting of interface in larger size droplet by the multiple minima present in the "mosaic structure" of supercooled liquids. The relation given by Eq. (2) was suggested to be valid only if the size of the droplet is much smaller than the mosaic elements.

In this study, we have considered a realistic system of an interface between two coexisting stable phases wetted by multiple intermediate metastable phases and we aim to obtain a scaling relation, if any, of the surface tension with the number of metastable phases(N) who contribute in wetting effect. For some particular energetics of the MS phases in different geometry we recover the scaling relation with radius, R (Eq. (2)), given above.

Surface energy between two phases can be determined using classical density functional theory (DFT)[12] considering the free energy density of different phases of an inhomogeneous system as a sum of free energy density of homogeneous medium and the spatial variation of density

$$\Omega_i[\rho(\mathrm{r})] = \int d\mathbf{r} \left[ f_{i,0}[\rho(\mathbf{r})] + \kappa \left( \nabla \rho(\mathbf{r}) \right)^2 \right] \tag{3}$$

Here $f_{i,0}[\rho(\mathrm{r})]$ is Landau (or, Helmholtz) free energy density function of the number density, $\rho(\mathrm{r})$ of the $i^{\text{th}}$ phase. $\kappa$ is related to the correlation length. One obtains the density profile of the inhomogeneous system by minimizing the free energy profile ($\delta \Omega[\rho(z)]/\delta \rho(z) = 0$; for a flat interface at xy-plane) and finally, surface tension is obtained as the extra free energy cost for the formation of interface $\gamma_{\text{M–SS}} = \left( \Omega[\rho(z)] - \Omega_{\text{M/SS}} \right)/A$, where A is the area of the interface.

In a previous work, surface tension between melt and stable solid phase has been determined in the presence of one metastable phase using this formalism[12]. However, in this work we aim to consider a more complicated and **predetermined** free energy surface in the presence of multiple metastable phases with different energetics, as shown in **Figure 1**. Therefore we have used a different theoretical formalism here.

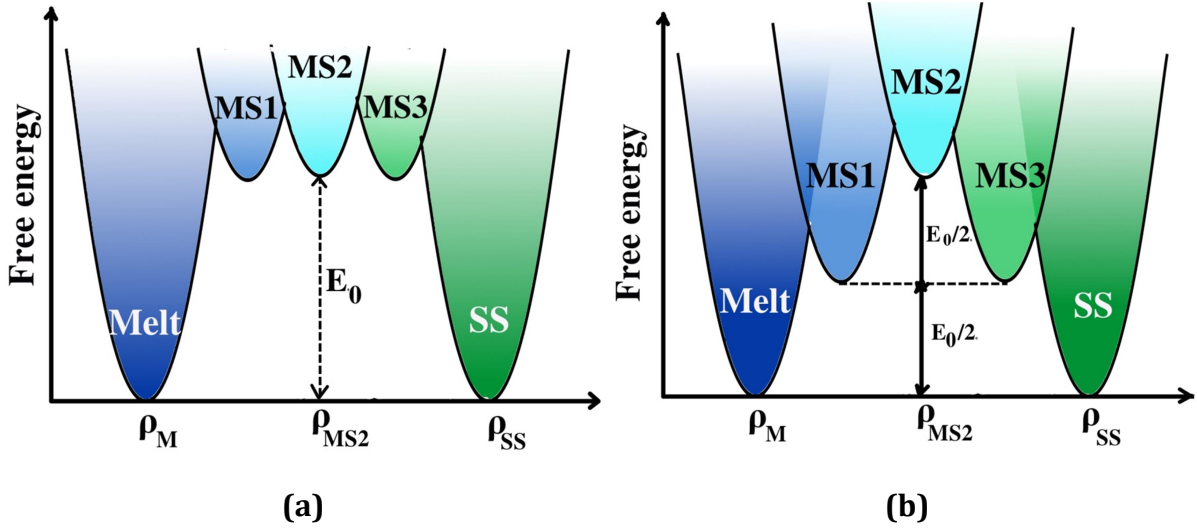

*Figure 1: Model systems of interface between stable solid and melt phase with three metastable phases. Two stable phases are taken to be at coexistence to calculate surface tension between them. (a) Model I consists of MS phases with similar energy, (b) Model II: MS phases with different energy arranged in a ladder like structure. The highest energy of the MS phase of Model II ($E_0$) is equal to the energy of MS phases in Model I.*

Cahn-Hilliard theory, although in a phenomenological description, allows one to calculate surface tension of an inhomogeneous system. Minimizing the similar free energy profile (Eq. (3)) with suitable boundary conditions, Cahn-Hilliard derived an analytical expression for the surface tension as $\gamma = 2 \int_{\rho_M}^{\rho_{SS}} [\kappa \Omega[\rho]]^{\frac{1}{2}} d\rho$.

Here $\rho_M$ and $\rho_{SS}$ are the equilibrium density of melt and stable solid phase respectively and $\kappa$ is related to the correlation length. Our goal is to determine the surface energy between melt and stable solid phase in the two model systems having different architecture of free energy surfaces of the metastable phases (**Figure 1**). The thermodynamic criteria to define surface tension between two phases are they have to be at coexistence with each other, i.e. in equilibrium with each other. Therefore, melt and stable solid (SS) phase are considered to be at coexistence in both the model systems. In Model I, all the metastable phases are considered to have similar free energy minima (**Figure 1(a)**), whereas in Model II, all the free energy minima of the metastable phases are arranged in a ladder like structure (**Figure 1(b)**).

Now, the surface energy of these model systems with three metastable phases can be written as a sum of eight terms

$$\gamma = \gamma_1 + \gamma_2 + \gamma_3 + \gamma_4 + \gamma_5 + \gamma_6 + \gamma_7 + \gamma_8$$

$$= 2\sqrt{\kappa} \left[ \sqrt{\frac{\lambda_M}{2}} \int_{\rho_M}^{\rho_1} (\rho - \rho_M) d\rho + \int_{\rho_1}^{\rho_{MS1}} \left[ \frac{\lambda_{MS1}}{2}(\rho - \rho_{MS1})^2 + E \right]^{1/2} d\rho + \int_{\rho_{MS1}}^{\rho_2} \left[ \frac{\lambda_{MS1}}{2}(\rho - \rho_{MS1})^2 + E \right]^{1/2} d\rho \right.$$
$$+ \int_{\rho_2}^{\rho_{MS2}} \left[ \frac{\lambda_{MS2}}{2}(\rho - \rho_{MS2})^2 + E \right]^{1/2} d\rho + \int_{\rho_{MS2}}^{\rho_3} \left[ \frac{\lambda_{MS2}}{2}(\rho - \rho_{MS2})^2 + E \right]^{1/2} d\rho$$
$$\left. + \int_{\rho_3}^{\rho_{MS3}} \left[ \frac{\lambda_{MS3}}{2}(\rho - \rho_{MS3})^2 + E \right]^{1/2} d\rho + \int_{\rho_{MS3}}^{\rho_4} \left[ \frac{\lambda_{MS3}}{2}(\rho - \rho_{MS3})^2 + E \right]^{1/2} d\rho + \sqrt{\frac{\lambda_{SS}}{2}} \int_{\rho_4}^{\rho_{SS}} (\rho - \rho_{SS}) d\rho \right]$$

(4)

Here $\lambda_M$, $\lambda_{MSi}$ and $\lambda_{SS}$ are the curvatures of the free energy surfaces of different phases. E is the free energy minima of metastable phases that can vary in the two model systems.

## A. Numerical results of surface tension between two stable phases in terms of multiple metastable phases (curvatures of all FES are equal)

In **Figure 1**, we have shown only three MS phases. However, in the numerical work, we have calculated surface energy between two stable phases in the presence of N number of metastable phase and we have varied N upto 10. $E_0$ is the energy of most metastable phase for both the model systems as shown in **Figure 1** and this parameter can be varied up to a higher limit of energy depending on the surfaces of melt and SS to have the contribution of metastable phases in reducing the surface tension between melt and SS.

We now present a scaling analysis of the problem. If we consider no intermediate phase between two stable forms, Cahn-Hilliard equation gives the expression for surface tension in the system as $\gamma_{M/SS} = \frac{\sqrt{2\kappa\lambda}}{4}(\rho_{SS} - \rho_M)^2$. Let us now consider N number of metastable phases between M and SS and all of them are at coexistence with M and SS and all of them have the same curvatures (**Figure 1a**). For such a system, we can minimize the free energy to obtain an expression of the total surface energy, given as $\gamma_{M/SS}^w = \sqrt{2\kappa\lambda}\frac{(\rho_{SS} - \rho_M)^2}{4} \times \frac{1}{(N+1)}$.

The comparison of the above two equations gives the relation between the surface tension between two stable phases (melt and stable solid) wetted by intermediate metastable phases ($\gamma_{M/SS}^w$) with the same in the absence of wetting ($\gamma_{M/SS}$) as $\gamma_{M/SS}^w = \frac{\gamma_{M/SS}}{N+1}$.

The simplicity of this 1/(N+1) dependence arises from the simple, although unphysical, arrangement of the metastable minima along the order parameter plane. In real world, the arrangement of the minima is bound to be more complex. We next consider a more complex and more realistic arrangement as shown in **Figure 1(b)**. If free energy minima of metastable phases are progressively destabilized, we could not solve for the final expression analytically. We carried out extensive calculations by varying N and find numerically (**Figure 2**) that the scaling relation of surface tension with the number of metastable phases (N) modifies into the following form where we introduce the exponent α.

$$\gamma_{M/SS}^{w} = \frac{\gamma_{M/SS}}{(N+1)^{\alpha}} \tag{5}$$

**Figure 1(a-b)** presents results on the surface free energy between melt and stable solid phase as a function of number of metastable phases (N) at different $E_0$ values. Unlike the result of Villain and Kirkpatrick *et al.*, here we observe that surface tension saturates after a number of MS phases for both the model systems except for model I at higher $E_0$ due to the artifact of the model. However, none of the curves can be fitted by the equation $\gamma_{M/SS}^{w} = \gamma_{M/SS}^{No-wetting}/N+1$ but by $\gamma_{M/SS}^{w} = \gamma_{M/SS}^{No-wetting}/(N+1)^{\alpha}$.

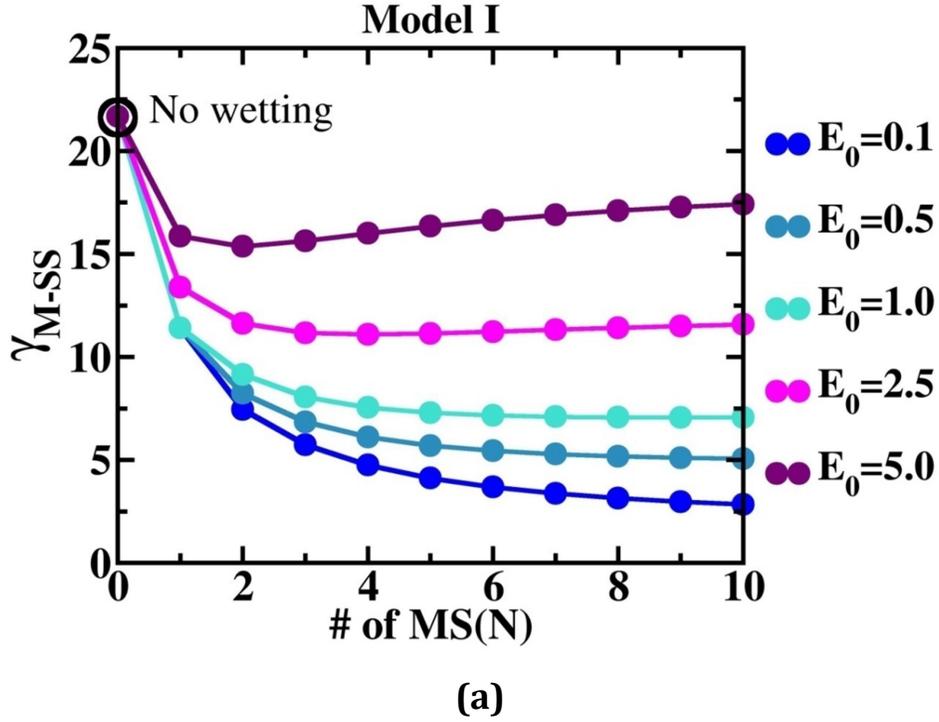

(a)

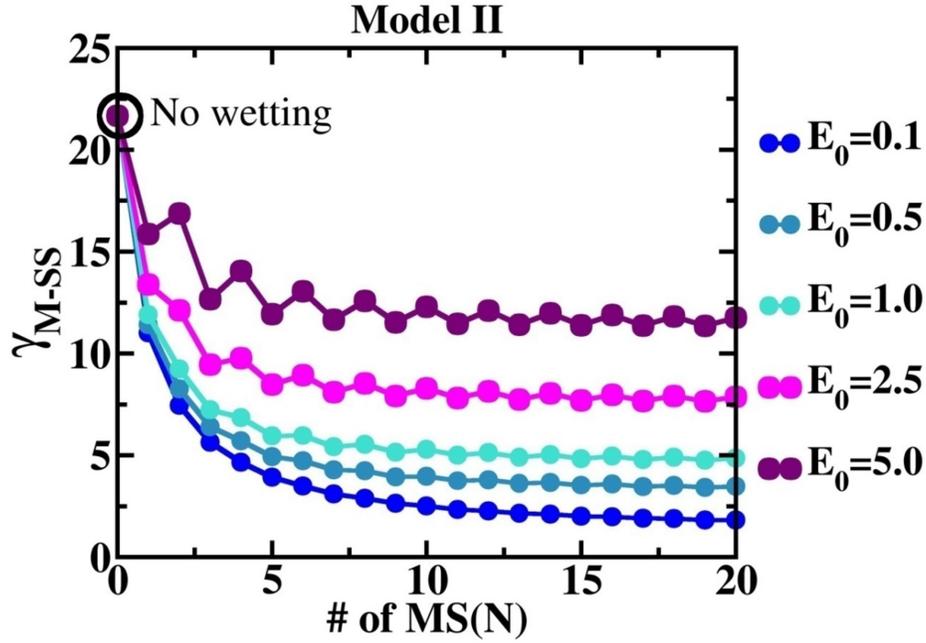

(b)

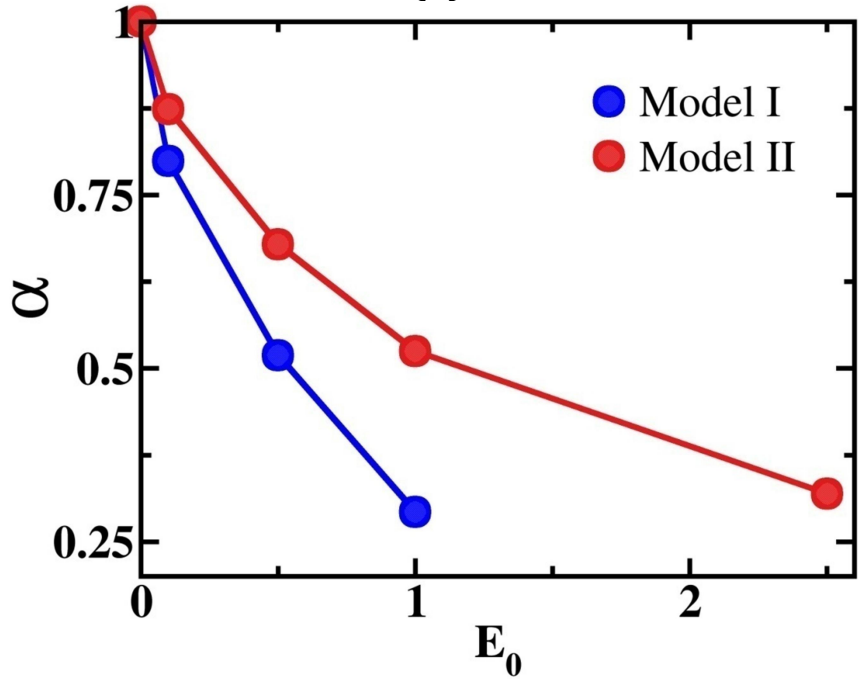

(c)

*Figure 2: Surface free energy of melt and stable solid phase($\gamma_{M/SS}$) in presence of multiple metastable phases for (a) Model I and (b)Model II. At the higher $E_0$ value, the increase of $\gamma_{M/SS}$ in model I is due to the artifact of the model; FES of MS phases invade into the stable phases resulting in an increase of $\gamma_{M/SS}$. (c) The scaling parameter α as a function of the energetics of the metastable phases ($E_0$ values) for Model I and Model II.*

***Different $\gamma^w_{M/SS}(N)$ curves are fitted to Eq. (5) to obtain α corresponding to different $E_0$ values. When $E_0$ is 0.0 that is all the phases are at coexistence, surface energy obeys $1/(N+1)$ dependence and with the increase of $E_0$, the dependence of surface energy on number of metastable phases (N) becomes weaker.***

The scaling relation depends mainly on two variables: (i) the energy of the metastable phases and (ii) the curvature of the free energy surface. If we consider the curvatures of free energy surfaces of all stable and metastable phases are to be equal, we obtain the dependence of the exponent on the energy $E_0$ as shown in **Figure 2(c)**. When $E_0$ is 0.0, all the metastable phases are at coexistence with the stable phases and we of course find α=1 as it obeys the relation of 1/N+1 derived earlier. As we pack more and more metastable phase with higher $E_0$ values, it shows weaker dependence on N for both the model systems and α decreases monotonically. For two model systems, the variation of α with $E_0$ is quite different and the scaling becomes $\sim(N+1)^{1/2}$ at $E_0$=0.5 for Model I and $E_0$=1.0 for Model II.

To connect these results on N-dependence of surface tension (N: number of MS phases) with that of fractional R-dependence result of random field Ising model by Villain (Eq. (2)), we need to establish the relation between the radius of nucleus (R) and the number of metastable phases (N)(N=f(R)) wetting the surface of nucleus which is not well defined and can vary from system to system and with different physiological condition. In a recent study of mineralogy, different mineral phases have been reported to coexist in different structures, sometime mosaic like and sometimes one surrounding another in a circular disk-like structure[30]. In our model of metastable phases waiting an interface, each phase would occupy a region of width of at least one molecular diameter. Therefore, the radius R should obey the condition $R = b\sigma N$, where b is a numerical constant of order unity and $\sigma$ is a molecular diameter. This ansatz gives rise to the radius dependence of the surface tension suggested by Villain using RFIM.

We have shown a schematic picture of the wetting of the surface of nucleus by various metastable phases in **Figure 3**. As the size (R) of the nucleus increases with time during nucleation process, more number of metastable phases can wet the surface of it and causes the reduction in surface energy contribution to the total nucleation free energy (Eq. (1)). The contribution of metastable phases in reducing the surface tension between the coexisting melt and SS phase ($\gamma^w_{M/SS}$) depends solely on the energy of the phase and spatiality. However, if the radius of the phase is larger, the number of different metastable phases surrounding the growing domain increases that leads to a higher probability of having such metastable phases that contribute in the reduction of $\gamma^w_{M/SS}$ significantly.

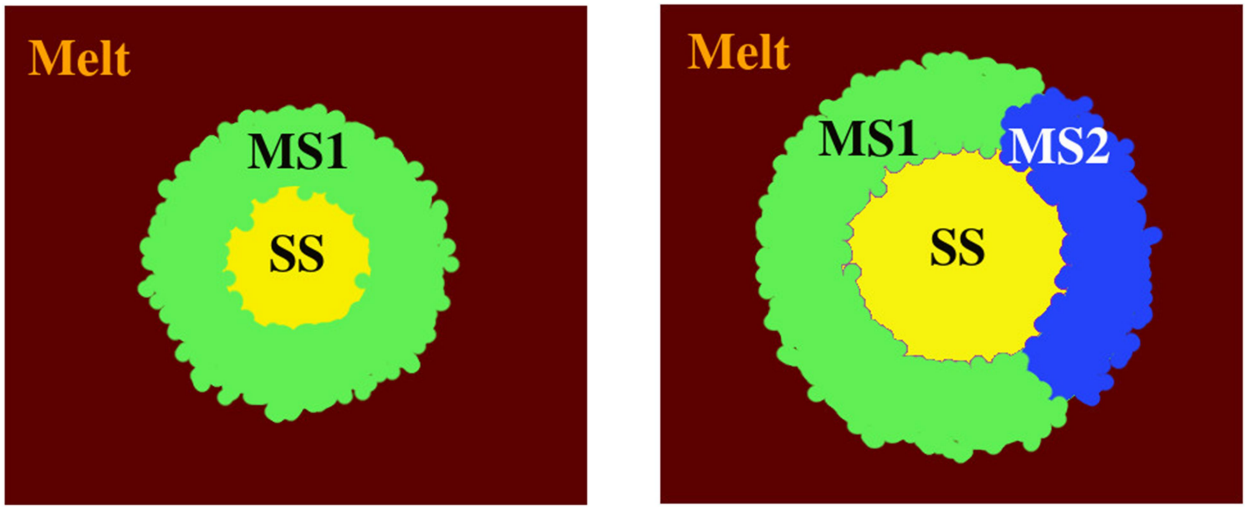

*Figure 3: Schematic picture of the wetting of surface of the stable solid phase by different number of metastable phases as it continues growing.*

## B. Temperature dependence of fractional dependence of surface tension on number of metastable phases (N)

Next, we want to investigate the scaling relation of surface tension with number of metastable phases when temperature changes which induces changes in various parameters, most glaring one is the curvatures of free energy surfaces. We assume that the change in the radius of curvature of stable phases is negligible compared to the change in that of metastable phases. Therefore, for numerical calculations, we fixed the curvatures of melt phase ($\lambda_M$) and stable solid phase ($\lambda_{SS}$) and we have varied $\lambda_{MS}$. At lower temperatures, we used a free energy surface of MS phase that is narrower than that of the melt phase and as the temperature increases, the ratio of $\lambda_{MS}/\lambda_M$ is allowed to approach 1.

For both the model systems, as the ratio of curvature approaches 1 at higher temperature, the effective surface tension between melt and stable solid phases decreases for different number of metastable phases that exist between them (**Figure 4(a-b)**). In these two figures, we have shown the situation for a particular $E_0(=1.0)$. We have computed the fitting parameter, α for different $E_0$ for both the model systems, shown in **Figure 4(c)**.

A new result of the present study is the strong dependence of the exponent α of the scaling relation between the surface tension and the number of metastable phases on the curvatures of the free energy surface. *As temperature increases that is the ratio of curvatures approaches 1, the fitting parameter, α increases suggesting stronger dependence on the number of metastable phases*. This results in more reduction of the surface energy between melt and SS phase at higher temperature in the presence of multiple metastable phases which helps in the nucleation of most stable solid state at higher temperature. This is a more realistic model system than the previous with all the free energy surfaces with similar curvatures. And this also recovers the scaling relation suggested by renormalisation group approach (Eq. (2)) for particular values of $E_0$ and $\lambda_{MS}$.

Therefore, we conclude that the fractional size dependence is not universal. Nevertheless, this general free energy surface based calculation recovers the result suggested by Villain and used by Xia-Wolynes to obtain Adam-Gibbs relation from the RFOT theory of glass transition.

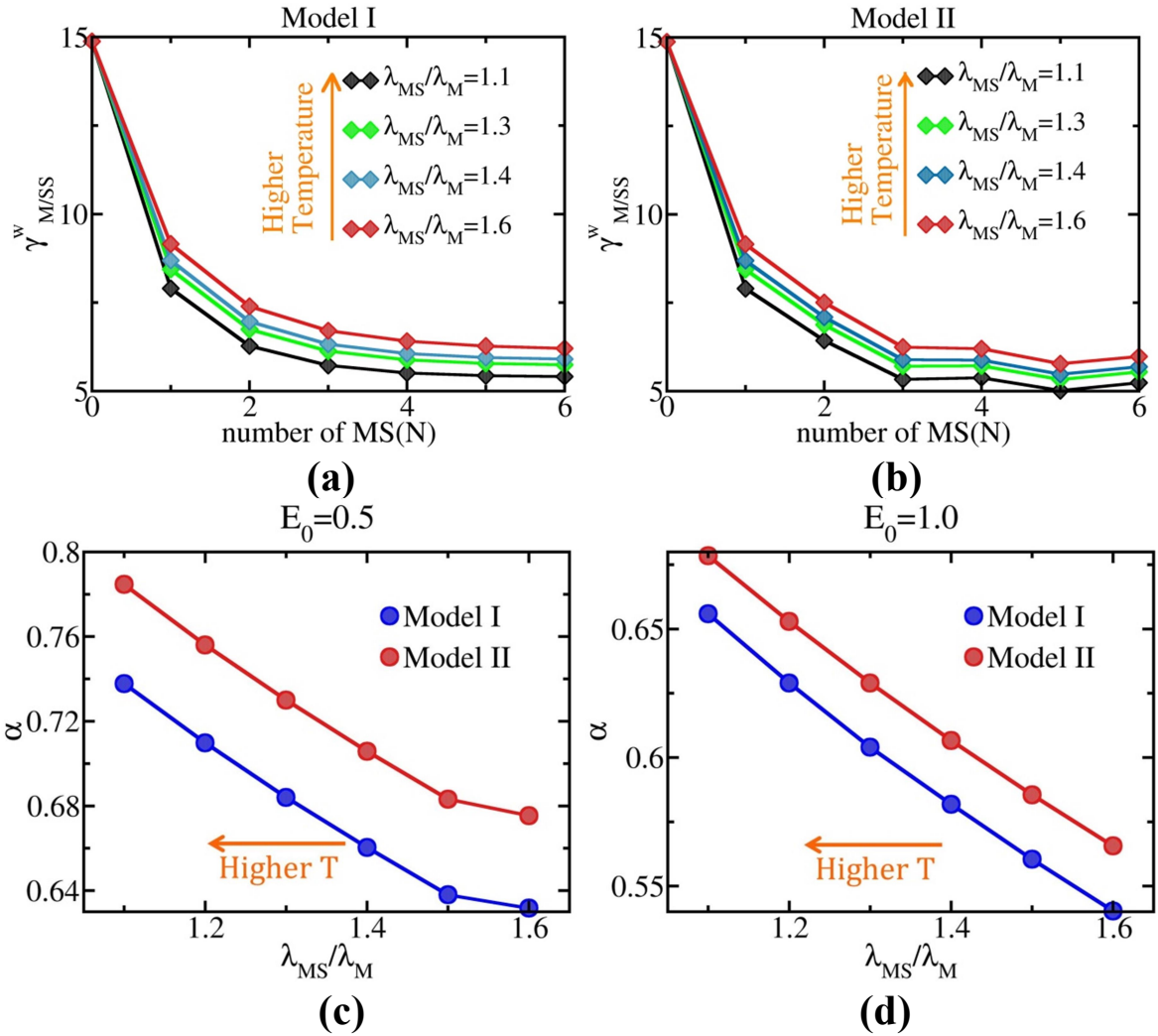

*Figure 4: Curvature of the free energy surfaces determines surface tension. This dependence of surface energy between melt and stable solid phases as a function of number of metastable phases (N) at different ratio of curvatures but same $E_0(=1.0)$ for (a) model I and (b) model II, (c) the scaling parameter, α as a function of the ratio of the curvatures for (c) $E_0=0.5$ and (d) $E_0=1.0$.*

In summary we have investigated the dependence of the surface tension between two stable phases on the number metastable phases wetting the surface of the growing nucleus of stable phase. This problem is related to a wide field of research consists of the nucleation of polymorphic solids, synthesis of a nanomaterial in the presence of other metastable phases of different materials etc.

The present study suggests the following scenario of phase transformation in the nanoscopic world. The initial phase (say, MS1) that forms from the parent melt phase fosters growth of the next, more stable than the initial phase but still metastable (MS2) phase with respect to the final phase. The kinetics of such processes is interesting, although substantially more complex. The rate equation of such processes can be written as

$$\frac{\partial P_M}{\partial t} = -k_{M \to MS1} P_M + k_{MS1 \to M} P_{MS1}$$

$$\frac{\partial P_{MS1}}{\partial t} = -k_{MS1 \to MS2} P_{MS1} - k_{MS1 \to M} P_{MS1} + k_{MS2 \to MS1} P_{MS2} + k_{M \to MS1} P_M - k_P P_{MS1} \quad (6)$$

$$\vdots$$

Here the rate constants ($k$) may be obtained from CNT but with the surface tension between two metastable phases (in coexistence with the same free energy), and the free energy gap. It would be interesting to explore the interplay between free energy stabilization and surface tension in the effective nucleation and even formation of a particular state.

Here we have considered different systems and conditions to investigate the scaling relationship of surface tension with number of metastable phases (N) and recovered the fractional radius dependence of surface tension suggested by Villain using renormalization group approach in the random field Ising model. We established this result for different systems with different energy condition and curvatures of metastable phases. The two points considered here are: (i) volume of each wetting phase and (ii) the arrangement of them. The minimum free energy requirement decides the arrangement placing similar Q's (Q: a suitable order parameter) next to each other, as envisaged in Ostwald's Step Rule. Villain's method using random field Ising model might allow more entropy because of many arrangement possibilities.

Our treatment is different from homogeneous DFT that assumes continuous decrease/increase of order parameter Q. The effect appears to be nearly the same as far as the decrease in the surface tension, but dramatically different in the prediction of the formation of metastable phases and also increase in the width of the interface. The ruggedness of the underlying free energy enforces certain width. Each minimum can contain a bit of the metastable phase and the width must be at least one molecular diameter per MS. In complex solids, it should be more.

In full DFT calculation with proper direct correlation function, the width of interface is calculated self-consistently with the surface tension. However, the over-all results should not be too different between the full DFT and Cahn-Hilliard approach. In the case of complex solids, the presence of rugged energy landscape within minima is essentially signature of an inhomogeneous liquid. It characterizes the energy landscape with sluggish relaxation and large non-Gaussian parameter which are characteristics of a glassy liquid.

**Acknowledgement**

The work was supported partly by Department of Science and Technology (DST), Govt. of India, Sir J. C. Bose fellowship, and Council of Scientific and Industrial Research (CSIR), India.